\documentclass[aps,showpacs,twocolumn,amsmath,amssymb,superscriptaddress,prl]{revtex4-1}

\usepackage[dvipdfmx]{graphicx}
\usepackage[dvipdfmx]{color}

\usepackage{bm}

\begin{document}

\title{Extrinsic-Intrinsic Crossover of the Spin Hall Effect Induced by Alloying}

\author{Akira Musha} \affiliation{Department of Applied Physics and Physico-Informatics, Keio University, Yokohama 223-8522, Japan}

\author{Yusuke Kanno}
\affiliation{Department of Applied Physics and Physico-Informatics, Keio University, Yokohama 223-8522, Japan}

\author{Kazuya Ando}
\email{ando@appi.keio.ac.jp}
\affiliation{Department of Applied Physics and Physico-Informatics, Keio University, Yokohama 223-8522, Japan}

\date{\today}

\begin{abstract}
We report the observation of the crossover between the extrinsic and intrinsic spin Hall effect induced by alloying. We found that the spin Hall angle, the ratio of the spin Hall conductivity to the electric conductivity, changes drastically by tuning the composition of Au-Cu alloy. The spin Hall angle changes the sign only in a limited range of the Cu concentration due to the extrinsic skew scattering, while the intrinsic contribution becomes dominant with increasing the Cu concentration. This observation provides essential information for fundamental understanding of spin-orbit physics. 
\end{abstract}

\maketitle

The spin Hall effect (SHE) is an emergent phenomenon arising from spin-orbit coupling (SOC), which was theoretically predicted about half a century ago~\cite{dyakonov1971current,Hirsch,Sinova,Murakami}. Since its first observation, this phenomenon has played a crucial role in exploring spin physics in condensed matter~\cite{Kato,Wunderlich,Valenzuela,Saitoh,KimuraPRL,hoffmann2013spin,RevModPhys.87.1213}. Specifically, the SHE enables electric manipulation of magnetization through spin-orbit torques, opening the field of spin-orbitronics~\cite{AndoPRL,liu2011spin,liu2012spinScience}. The inverse process of the SHE allows electric detection of spin currents, enabling to reveal a variety of phenomena, such as the spin Seebeck effect and spin pumping from magnetic insulators~\cite{kajiwara,seebeckinsulator}.

The SHE shares the same origin as the anomalous Hall effect (AHE): both effects involve the generation of a transverse spin current from an applied charge current through the SOC~\cite{nagaosa2010anomalous,RevModPhys.87.1213}. The spin-current generation arises from intrinsic and extrinsic mechanisms. The intrinsic contribution originates from the Berry curvature associated with the Fermi surface and the band structure of the material~\cite{karplus1954hall,Sinova,Murakami,tanaka2008intrinsic}, whereas the extrinsic mechanism, including the skew scattering and side jump, originates from the spin-dependent scattering on structural defects or impurities~\cite{smit1958spontaneous,berger1970side}. The power law dependence of the anomalous Hall conductivity on the longitudinal conductivity changes in various conductivity regimes; the extrinsic skew scattering mechanism appears in the clean limit, whereas the intrinsic contribution is dominant in films with low electric conductivity~\cite{nagaosa2010anomalous}. The extrinsic-intrinsic crossover, an important feature of the AHE, has been observed in a wide range of ferromagnets, providing a fundamental understanding of the spin-transport physics~\cite{nagaosa2010anomalous}. In contrast, despite the similarities between the AHE and SHE, the crossover between the extrinsic and intrinsic regimes of the SHE was observed only recently in Pt, the prototypical SHE metal, by tuning the conductivity~\cite{sagasta2016tuning}.

The extrinsic contributions to the SHE can be manipulated by alloying: by changing the combination of the host and impurity metals or by changing the concentration of the impurities~\cite{gradhand2010extrinsic,obstbaum2016tuning}. The engineering of the SHE by alloying has been reported previously for a variety of systems, including Cu, Pt, and Au-based alloys~\cite{niimi2012giant,niimi2011extrinsic,niimi2014extrinsic,chen2017tunable,nguyen2016enhanced,ramaswamy2017extrinsic,zhu2018highly,qu2018inverse,laczkowski2017large,laczkowski2014experimental,wen2017temperature,wu2016spin,zou2016large}. These studies have demonstrated that metal alloying opens a promising route to tune the SHE in a simple and robust way. However, despite the significant efforts on the study of the SHE, the crossover between the extrinsic and intrinsic regimes induced by tuning alloying has remained elusive; the SHE in metallic alloys has been shown to be governed by either the extrinsic mechanism (e.g. Cu(Bi), Cu-Pt, Au-Ta)~\cite{niimi2012giant,ramaswamy2017extrinsic,laczkowski2017large} or the intrinsic mechanism (e.g. Pt-Al, Pt-Au, Au-W)~\cite{laczkowski2017large,nguyen2016enhanced,zhu2018highly}.

In this Letter, we report the observation of the crossover of the spin Hall effect between the two distinct regimes, the extrinsic impurity scattering and intrinsic Berry curvature mechanism, induced by tuning the composition of Au-Cu alloy. The Au-Cu alloy forms a solid solution in the full composition range, making it a suitable system for studying the effect of alloying on the SHE~\cite{wen2017temperature,wu2016spin,zou2016large}. We show that the effective spin Hall angle (SHA) of the $\rm Au_{100-\it x}Cu_{\it x}$ alloy changes drastically with the Cu concentration $x$. We found that the sign of the effective SHA changes from positive to negative by changing $x$ from 0 to 5. This sign reversal is consistent with an $ab$ $initio$ calculation of the SHE due to the skew scattering in Au with dilute Cu impurities~\cite{gradhand2010spin}. Furthermore, by further increasing $x$, the effective SHA increases linearly and changes the sign from negative to positive at $x\sim 16$. These results provide evidence of the crossover of the SHE from the extrinsic regime, where the skew scattering is dominant, to the intrinsic regime in the metallic alloy.

\begin{figure}[tb]
\includegraphics[scale=1]{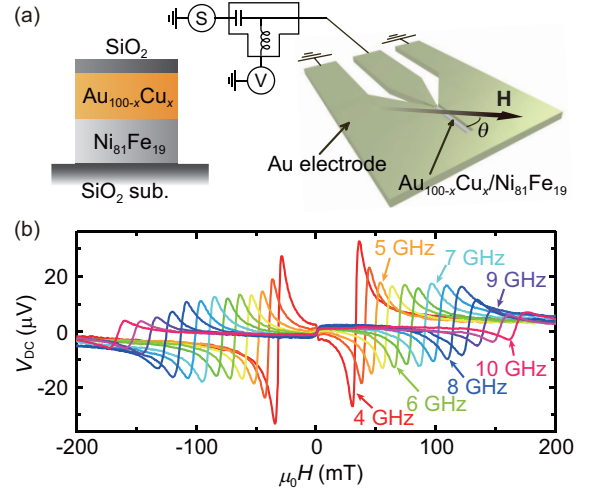}
\caption{
(a) Schematic illustration of the SiO$_{2}$/Au$_{100-\it x}$Cu$_{\it x}$/Ni$_{81}$Fe$_{19}$ film and the experimental setup for the ST-FMR measurement. (b) Magnetic field $H$ dependence of DC voltage $V_\text{DC}$ for the Au$_{95.1}$Cu$_{4.9}$/Ni$_{81}$Fe$_{19}$ bilayer measured with the RF current frequency from $f=4$ to 10 GHz and the power of $P=100\;\rm mW$. The external magnetic field was applied at the angle of $\theta=225^{\circ}$.
}
\label{fig1} 
\end{figure}

We measured the SHE of the $\rm Au_{100-\it x}Cu_{\it x}$ alloy using the spin-torque ferromagnetic resonance (ST-FMR) for SiO$_{2}$(4)/Au$_{100-\it x}$Cu$_{\it x}$(8)/Ni$_{81}$Fe$_{19}$(8)/SiO$_2$-substrate films, where the numbers in parentheses represent the thickness in the unit of $\rm nm$ [see Fig.~\ref{fig1}(a)]~\cite{liu2011spin}. The films were deposited on thermally oxidized $\rm Si$ substrates by RF magnetron sputtering in the based pressure around $1 \times 10^{-5}$ Pa. The $\rm Au_{100-\it x}Cu_{\it x}$ layer was co-sputtered with rotating the substrate. The composition ratio was varied by changing the Cu sputtering power between 0 to 20 W, while the Au sputtering power was fixed at 50 $\rm W$. The Cu concentration was determined by the X-ray fluorescence analysis. To measure the ST-FMR for the Au$_{100-\it x}$Cu$_{\it x}$/Ni$_{81}$Fe$_{19}$ bilayer, the film was patterned into a $10\:\rm\mu m \times 150\:\rm\mu m$ rectangle shape using photolithography. On the edges of the Au$_{100-\it x}$Cu$_{\it x}$/Ni$_{81}$Fe$_{19}$ bilayer, Au/Ti electrodes were attached using the photolithography and sputtering, as shown in Fig.~\ref{fig1}(a).

For the ST-FMR measurement, an RF charge current was applied along the longitudinal direction of the Au$_{100-\it x}$Cu$_{\it x}$/Ni$_{81}$Fe$_{19}$ bilayer using a signal generator, and an in-plane external magnetic field ${\bf H}$ was applied with an angle of $\theta$ from the longitudinal direction of the device, as shown in Fig.~\ref{fig1}(a). In the bilayer, the RF charge current generates a transverse spin current through the SHE in the Au$_{100-\it x}$Cu$_{\it x}$ layer. The transverse spin current injected into the Ni$_{81}$Fe$_{19}$ layer, as well as an Oersted field due to the current flow in the Au$_{100-\it x}$Cu$_{\it x}$ layer, exerts in-plane and out-of-plane torques on the magnetization in the Ni$_{81}$Fe$_{19}$ layer, leading to magnetization precession. The magnetization precessing induces an oscillation of the resistance of the bilayer through the anisotropic magnetoresistance (AMR). Due to the mixing of the applied RF current and oscillating resistance, DC voltage is generated in the bilayer under the FMR condition. We measured the DC voltage $V_\text{DC}$ for the Au$_{100-\it x}$Cu$_{\it x}$/Ni$_{81}$Fe$_{19}$ bilayer using a bias tee as shown in Fig.~\ref{fig1}(a). From the spectral shape of the $V_\text{DC}$ signal, the in-plane and out-of-plane torques can be separated, since the phase of these torques differs by $\pi/2$, which results in symmetric and antisymmetric shapes of $V_\text{DC}$. The DC voltage due to the sum of the in-plane and out-of-plane torques is expressed as~\cite{fang2011spin},
\begin{align}
	V_{\rm DC}&=V_{\rm sym}\frac{W^2}{\left(\mu_{0}H-\mu_{0}H_{\rm res}\right)^2+W^2}\nonumber\\
	&+V_{\rm antisym}\frac{W\left(\mu_{0}H-\mu_{0}H_{\rm res}\right)}{\left(\mu_{0}H-\mu_{0}H_{\rm res}\right)^2+W^2}, \label{ST-FMR}
\end{align}
where $V_{\rm sym}$ and $V_{\rm antisym}$ represent the magnitude of the symmetric and antisymmetric voltage:
\begin{align}
	V_{\rm sym}=\frac{\Delta RI}{2}\mu_{0}h_{ z}\sin2\theta \frac{\mu_{0}H_{\rm res}+\mu_{0}M_{\rm s}}{W(2\mu_{0}H_{\rm res}+\mu_{0}M_{\rm s})}\frac{1}{\sqrt{1+\displaystyle \frac{\mu_{0}M_{\rm s}}{\mu_{0}H_{\rm res}}}}, \label{V_sym}
\end{align}
\begin{align}
	V_{\rm antisym}=\frac{\Delta RI}{2}\mu_{0}h_{ y} \sin2\theta \cos\theta \frac{\mu_{0}H_{\rm res}+\mu_{0}M_{\rm s}}{W(2\mu_{0}H_{\rm res}+\mu_{0}M_{\rm s})}. \label{V_antisym}
\end{align}
Here, $\Delta R$ is the resistance change of the bilayer due to the AMR, $I$ is the amplitude of the longitudinal RF current, $\mu_{0}M_{\rm s}$ is the saturation magnetization, $W$ is the FMR linewidth, and $H_{\rm res}$ is the FMR field. As shown in Eqs.~(\ref{V_sym}) and (\ref{V_antisym}), the out-of-plane effective field $h_{z}$ due to a dampinglike torque induced by the SHE generates the symmetric voltage $V_{\rm sym}$, while the in-plane effective field $h_{y}$, dominated by the Oersted field, generates the antisymmetric voltage $V_{\rm antisym}$. We measured the mixing voltage  $V_\text{DC}$ for the Au$_{100-\it x}$Cu$_{\it x}$/Ni$_{81}$Fe$_{19}$ bilayers at the room temperature.

\begin{figure*}[bt]
\includegraphics[scale=1]{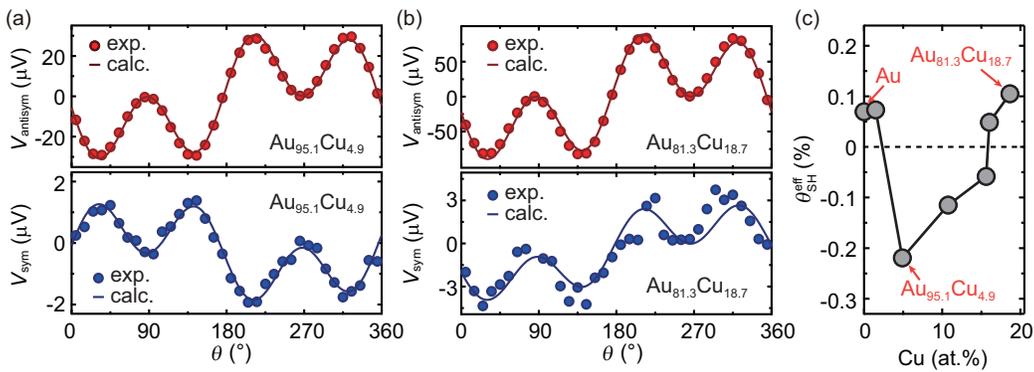}
\caption{
Magnetic field angle $\theta$ dependence of the antisymmetric $V_{\rm antisym}$ and symmetric $V_{\rm sym}$ components of the ST-FMR spectra for the Au$_{100-\it x}$Cu$_{\it x}$/Ni$_{81}$Fe$_{19}$ bilayer with (a) $x=4.9$ and (b) $x=18.7$. The solid circles are the experimental data and solid curves are the fitting result using a function proportional to $\sin 2\theta \cos \theta$. (c) Cu concentration $x$ dependence of the effective spin Hall angle $\theta^{\rm eff}_{\rm SH}$ for the Au$_{100-\it x}$Cu$_{\it x}$/Ni$_{81}$Fe$_{19}$ bilayer. 
}
\label{fig2} 
\end{figure*}

Figure~\ref{fig1}(b) shows the ST-FMR spectra for the Au$_{95.1}$Cu$_{4.9}$/Ni$_{81}$Fe$_{19}$ bilayer measured with the applied RF current frequency from $f=4$ to 10 GHz and power of $P=100 \rm \;mW$. For the measurement, the external magnetic field ${\bf H}$ was applied at $\theta=225^{\circ}$ [see also Fig.~\ref{fig1}(a)]. The resonance field of $V_\text{DC}$ changes systematically with $f$, which is in good agreement of Kittel formula. We also note that the sign of the $V_\text{DC}$ signals is reversed by reversing the magnetic field, consistent with the prediction of the ST-FMR~\cite{liu2011spin}.

In Fig.~\ref{fig2}(a), we show field-angle $\theta$ dependence of $V_{\rm sym}$ and $V_{\rm antisym}$, extracted from the $V_\text{DC}$ signals measured at $f=7$ GHz, for the Au$_{95.1}$Cu$_{4.9}$/Ni$_{81}$Fe$_{19}$ bilayer. Figure~\ref{fig2}(a) shows that $\theta$ dependence of $V_{\rm antisym}$ is entirely fitted by a function proportional to $\sin 2\theta \cos \theta$, indicating that $h_y$ is independent of $\theta$ [see Eq.~(\ref{V_antisym})]. This is consistent with the prediction that the $V_{\rm antisym}$ is dominated by the Oersted field created by the current flowing in the Au$_{95.1}$Cu$_{4.9}$ layer. In contrast, the $\theta$ dependence of $V_{\rm sym}$ indicates that $h_{z}\propto \cos\theta$; the $\theta$ dependence of $V_{\rm sym}$ is entirely fitted by a function proportional to $\sin 2\theta \cos \theta$ [see Eq.~(\ref{V_sym})]. The angular dependence of $h_{z}$ is consistent with the prediction of the spin-transfer mechanism, where a spin current generated by the SHE in the Au$_{95.1}$Cu$_{4.9}$ layer is absorbed in the Ni$_{81}$Fe$_{19}$ layer and exerts a dampinglike torque on the magnetization~\cite{gambardella2011current}. This confirms that the SHE in the Au$_{95.1}$Cu$_{4.9}$ layer is responsible for the $V_{\rm sym}$ signals. Here, the interface Rashba effect at a Au/Ni$_{81}$Fe$_{19}$ interface is known to be negligible~\cite{wen2017temperature,grytsyuk2016k}.

We found that the sign of $V_{\rm sym}$ is changed by changing the Cu concentration of the Au$_{100-\it x}$Cu$_{\it x}$ layer. In Fig.~\ref{fig2}(b), we show the field-angle $\theta$ dependence of $V_{\rm antisym}$ and $V_{\rm sym}$ for the Au$_{81.3}$Cu$_{18.7}$/Ni$_{81}$Fe$_{19}$ bilayer. We note that the sign of $V_{\rm sym}$ is reversed, while the sign of $V_{\rm antisym}$ is not changed, compared with the result for the Au$_{95.1}$Cu$_{4.9}$/Ni$_{81}$Fe$_{19}$ bilayer. This result indicates that the direction of $h_z$, or the dampinglike torque generated by the SHE in the Au$_{100-\it x}$Cu$_{\it x}$ layer, is opposite between $x=4.9$ and $x=18.7$; the sign of the SHA is reversed by changing the Cu concentration.

To investigate the sign reversal of the SHA in the Au$_{100-\it x}$Cu$_{\it x}$ film induced by changing the Cu concentration, we summarize the Cu-concentration dependence of the effective SHA, defined as 
\begin{align}
\theta^{\rm eff}_{\rm SH}=\frac{2e}{\hbar}\frac{\mu_{0}M_{\rm s}d_{\rm F}h_{z}}{j_{\rm c}} \label{effSHA}
\end{align}
in Fig.~\ref{fig2}(c), where $d_\text{F}$ is the thickness of the Ni$_{81}$Fe$_{19}$ layer. To obtain $\theta^{\rm eff}_{\rm SH}$, the out-of-plane dampinglike effective field $h_z$ was determined from $V_\text{sym}$ with measured values of $\Delta R$ for each bilayers with various $x$ [see Eq.~(\ref{effSHA})]. In Eq.~(\ref{effSHA}), $j_\text{c}$ is the current density flowing in the Au$_{100-\it x}$Cu$_{\it x}$ layer, which was estimated for each Au$_{100-\it x}$Cu$_{\it x}$/Ni$_{81}$Fe$_{19}$ bilayers with various $x$ by monitoring the current-induced resistance change due to the Joule heating~\cite{tshitoyan2015electrical}. In Fig.~\ref{fig2}(c), the positive sign of $\theta^{\rm eff}_{\rm SH}$ indicates that the sign of the SHA is same as that for Pt.

Figure~\ref{fig2}(c) demonstrates that the sign of $\theta^{\rm eff}_{\rm SH}$ changes two times by changing the Cu concentration of the Au$_{100-\it x}$Cu$_{\it x}$ layer from $x=0$ to 20. For the Au$_{100-\it x}$Cu$_{\it x}$ film with $x=0$, or the Au film, the sign of $\theta^{\rm eff}_{\rm SH}$ is positive. The sign, as well as the magnitude, of $\theta^{\rm eff}_{\rm SH}$ is consistent with the SHA for Au reported previously~\cite{isasa2015temperature,vlaminck2013dependence,an2018manipulation,mosendz2010quantifying}. Figure~\ref{fig2}(c) shows that with increasing the Cu concentration, $\theta^{\rm eff}_{\rm SH}$ changes drastically, and the sign is reversed at the Cu concentration of $x=4.9$. By further increasing the Cu concentration, $\theta^{\rm eff}_{\rm SH}$ increases gradually and changes the sign from negative to positive at the Cu concentration around $x=16$. The sign reversal of the SHA has been observed for Au-based alloy, such as Au-Ta alloy~\cite{qu2018inverse,laczkowski2017large}. In the Au-Ta alloy, the sign of the SHA changes from positive to negative with increasing the amount of Ta in Au. The sign reversal is reasonable because the sign of the SHA of the host metal Au and the impurity metal Ta is opposite. In this case, it is possible that the SHA in the alloy becomes zero at a certain concentration which depends on the relative strength of the SOC between the host and impurity metals. In contrast to this situation, in the Au-Cu alloy, the sign of the SHA of the impurity metal Cu is same as that of the host metal Au~\cite{ramaswamy2017extrinsic,an2016spin}, and thus the sign reversal of the SHA induced by increasing the Cu concentration is nontrivial.

\begin{figure}[tb]
\includegraphics[scale=1]{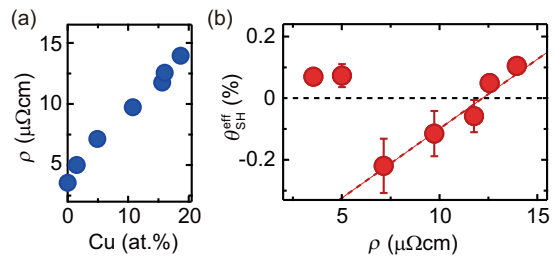}
\caption{
(a) The electrical resistivity $\rho$ of the Au$_{100-\it x}$Cu$_{\it x}$ film. (b) $\rho$ dependence of the effective spin Hall angle $\theta^{\rm eff}_{\rm SH}$. The dotted line in red is the linear fitting result for the measured values of $\theta^{\rm eff}_{\rm SH}$ above $\rho=7.15$ $\mu\Omega$cm.
}
\label{fig3} 
\end{figure}

The negative sign of $\theta^{\rm eff}_{\rm SH}$ can be attributed to the SHE due to the skew scattering in the Au$_{100-\it x}$Cu$_{\it x}$ alloy. In metallic alloys, generally, the intrinsic and side-jump contributions play an important role for high impurity concentrations, while the skew scattering is predominant in the dilute limit of impurity concentration~\cite{lowitzer2011extrinsic,fert2011spin}. For the Au-Cu alloy, the SHA due to the skew scattering in the dilute Cu limit was calculated based on the relativistic Korringa-Kohn-Rostoker method combined with the solution of a linearized Boltzmann equation~\cite{gradhand2010spin}. This calculation shows that the sign of the SHA is negative. Our experimental finding of the negative $\theta^{\rm eff}_{\rm SH}$, shown in Fig.~\ref{fig2}(c), is consistent with this calculation.

To further gain insight into the mechanism of the sign changes of the SHA, we plotted $\theta^{\rm eff}_{\rm SH}$ as a function of the electrical resistivity $\rho$ of the Au$_{100-\it x}$Cu$_{\it x}$ layer as shown in Figs.~\ref{fig3}(a) and \ref{fig3}(b). Figure~\ref{fig3}(b) shows that $\theta^{\rm eff}_{\rm SH}$ changes the sign at $\rho \simeq 5$ $\mu\Omega$cm and increases linearly with $\rho$ for $\rho > 5$ $\mu\Omega$cm. Since the SHA due to the skew scattering, which is responsible for the negative sign of $\theta^{\rm eff}_{\rm SH}$, is independent of the resistivity $\rho$, the linear increase of $\theta^{\rm eff}_{\rm SH}$ with $\rho$ can only be attributed to the side-jump or intrinsic mechanisms with positive $\theta^{\rm eff}_{\rm SH}$. Here, in contrast to the $\rho$-independent SHA of the skew scattering, the SHA due to the side jump and intrinsic mechanism is proportional to $\rho$; the SHA $\theta_\text{SH}$ is expressed as $\theta_\text{SH}=\theta^{\rm SS}_{\rm SH}+\sigma_{\rm SH}\rho$, where $\theta^{\rm SS}_{\rm SH}$ is the SHA of the skew scattering, and $\sigma_{\rm SH}$ is the spin Hall conductivity of the side jump and intrinsic mechanism~\cite{0034-4885-78-12-124501}.

We obtained the spin Hall conductivity $\sigma_{\rm SH}=475$ $\Omega^{-1}\rm cm^{-1}$ for the Au$_{100-\it x}$Cu$_{\it x}$ film from the linear fitting of the $\rho$ dependence of $\theta^{\rm eff}_{\rm SH}$, shown in Fig.~\ref{fig3}(b). This value is consistent with the spin Hall conductivity of Au due to the intrinsic mechanism, $\sigma_{\rm SH}=100\sim 800$ $\Omega^{-1}\rm cm^{-1}$~\cite{isasa2015temperature,gradhand2011calculating,guo2009ab,chadova2015separation}, indicating that the intrinsic mechanism, rather than the side jump, dominates the SHE in the Au$_{100-\it x}$Cu$_{\it x}$ alloy with $\rho >5$ $\mu\Omega$cm. Here, for the estimation of $\sigma_{\rm SH}$, we assumed that the spin-diffusion length, $\sim 5$ nm in Au-Cu alloy~\cite{wu2016spin,zou2016large}, does not change significantly with the Cu concentration in this $x$ range. We also neglected the spin memory loss at the interface~\cite{liu2014interface,rojas2014spin}, and thus the estimated spin Hall conductivity is the lower bound. We note that the sign of $\theta^{\rm eff}_{\rm SH}$, shown in Fig.~\ref{fig2}(c), indicates that the skew scattering with negative $\theta^{\rm eff}_{\rm SH}$ is dominant in the dilute regime ($x<16$), while the intrinsic mechanism with positive $\theta^{\rm eff}_{\rm SH}$ is dominant in the concentrated alloy regime ($x>16$). This result is consistent with a calculation based on the Kubo-St\v{r}eda linear-response formalism~\cite{lowitzer2011extrinsic}. The calculation shows that the intrinsic contribution of the effective medium dominates in concentrated alloys, whereas the skew scattering contribution shows in general a diverging behavior in the dilute alloy regime.

In conclusion, we demonstrated that the SHE in Au$_{100-\it x}$Cu$_{\it x}$ alloy changes drastically with the Cu concentration $x$. By changing the Cu concentration, we found that the sign of the effective SHA is reversed; the sign of the SHA becomes negative only in the range of $5< x < 16$, despite the positive SHA of pure Au and Cu. This finding is consistent with the $ab$ $initio$ calculation of the skew scattering that predicts the negative sign of the SHA in Au with in the dilute Cu impurities. Furthermore, we found the linear increase of the effective SHA with the resistivity of the Au$_{100-\it x}$Cu$_{\it x}$ alloy. These results demonstrate that the skew scattering with the negative SHA dominates the SHE in the Au$_{100-\it x}$Cu$_{\it x}$ alloy with $x\sim 5$, whereas the intrinsic contribution with the positive SHA to the SHE increases with the resistivity and is dominant when $x>16$, demonstrating an important correspondence between the SHE and AHE, the extrinsic-intrinsic crossover.

\begin{acknowledgments}
This work was supported by JSPS KAKENHI Grant Numbers 26220604, 26103004, the Asahi Glass Foundation, JGC-S Scholarship Foundation, and Spintronics Research Network of Japan (Spin-RNJ). 
\end{acknowledgments}

%\bibliography{ref.bib}


\end{document}